\documentclass[twocolumn,english,showpacs,preprintnumbers,epl]{revtex4-1}
\usepackage[latin9]{inputenc}
\usepackage{color}
\usepackage{amsmath}
\usepackage{amssymb}
\usepackage{graphicx}
\usepackage{esint}
\usepackage{ulem}
\makeatletter
\@ifundefined{textcolor}{}
{%
 \definecolor{BLACK}{gray}{0}
 \definecolor{WHITE}{gray}{1}
 \definecolor{RED}{rgb}{1,0,0}
 \definecolor{GREEN}{rgb}{0,1,0}
 \definecolor{BLUE}{rgb}{0,0,1}
 \definecolor{CYAN}{cmyk}{1,0,0,0}
 \definecolor{MAGENTA}{cmyk}{0,1,0,0}
 \definecolor{YELLOW}{cmyk}{0,0,1,0}
}

\usepackage{aecompl}

\usepackage{epsfig}\usepackage{dcolumn}\usepackage{bm}

\usepackage{babel}

\makeatother

\usepackage{babel}
\begin{document}

\title{
Effect of Inter-Adatoms Correlations on the Local Density of States of Graphene}

\author{A. C. Seridonio$^{1,2}$, K. Kristinsson$^{3}$, M. de Souza$^{1}$,
F. M. Souza$^{4}$, L. H. Guessi$^{1}$, R. S. Machado$^{2}$, and
I. A. Shelykh$^{3,5}$ }

\affiliation{$^{1}$
IGCE, Unesp - Univ Estadual Paulista, Departamento de F\'{i}sica, 
 13506-900, Rio Claro, SP, Brazil\\
 $^{2}$Departamento de F\'{i}sica e Qu\'{i}mica, Unesp - Univ Estadual
Paulista, 15385-000, Ilha Solteira, SP, Brazil\\
 $^{3}$Division of Physics and Applied Physics, Nanyang Technological
University 637371, Singapore\\
 $^{4}$Instituto de F\'{i}sica, Universidade Federal de Uberlândia,
38400-902, Uberlândia, MG, Brazil\\
 $^{5}$Science Institute, University of Iceland, Dunhagi-3, IS-107,
Reykjavik, Iceland }
\begin{abstract}
We discuss theoretically the local density of states  (LDOS) of a graphene sheet hosting
two distant adatoms located at the center of the hexagonal cells.
By putting  laterally a Scanning Tunneling Microscope (STM) tip over a carbon atom, two remarkable novel
effects can be detected:
i) a multilevel structure
in the LDOS and ii) beating patterns  in the induced
LDOS.
We show that both phenomena occur nearby the
Dirac points and are
highly anisotropic.
Furthermore, we propose 
conductance  experiments employing STM as a probe for the observation of such exotic manifestations in the LDOS of graphene induced by inter-adatoms correlations.
\end{abstract}

\pacs{72.80.Vp, 07.79.Cz, 72.10.Fk}

\maketitle

\textit{Introduction.-}
A graphene is a genuine two-dimensional (2D) monolayer system formed by carbon atoms packed into a hexagonal honeycomb lattice \cite{Novoselov,Peres,CNeto1}. A remarkable feature of such a system is the existence of Dirac cones at the corners of the Brillouin zone in its band structure, similar to those appearing in the relativistic dispersion of a massless particle. Consequently, graphene based systems provide appropriate conditions for emulation of relativistic phenomena in the domain of condensed matter physics. Interestingly enough, the appearance of quasi-relativistic massless Dirac fermions have been reported also in bulk molecular conductors \cite{Katayama} and topological insulators \cite{Hasan2010}. Recent experimental and theoretical works demonstrated the possibility of effective controllable adsorption of single magnetic impurities, the so-called adatoms, by an individual graphene sheet \cite{Eelbo1,Eelbo2,abInitio1}. To explore the physical properties of such adatoms as well as their effects on the properties of the host, Scanning Tunneling Microscope (STM) technique has been recognized as the most efficient experimental tool \cite{STMreview}. An STM setup consists of a metallic tip capable of detecting the local density of states (LDOS) via differential conductance
measurements.
\begin{figure}[!]
\includegraphics[width=0.5\textwidth,height=0.18\textheight]{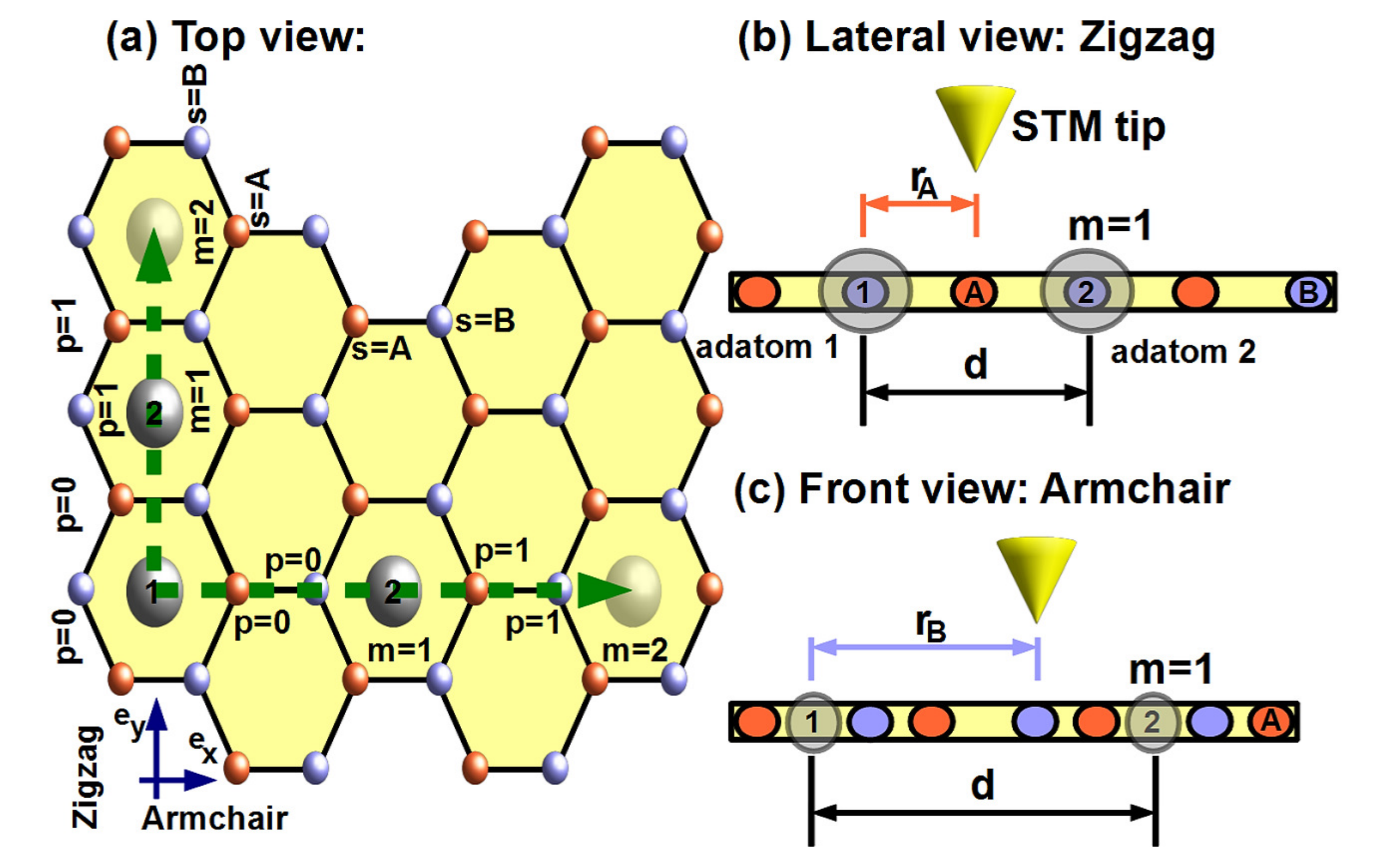}
\caption{\label{fig:Pic1} (Color online) (a) Two adatoms labeled by $1$ and
$2$ are placed far apart at the center of the hexagonal cells for
a given inter-adatoms distance ${\bold d}$ along the zigzag and armchair
directions. The shaded adatoms represent a larger separation between
the adatoms $1$ and $2$. In panels (b) and (c), the graphene LDOS
at ${\bold r}_{s}$ ($s=A,B$) can be probed by an STM tip 
in the zigzag and armchair directions.}
\end{figure}

Notably, the tip perceives a fascinating phenomenon involving
electronic scattering by impurities, known as Friedel oscillations,
which appears in the conductance signal as a damped oscillatory pattern
when the tip position is varied \cite{Friedel1,Friedel2}.
The properties of magnetic adatoms in graphene have been addressed theoretically
within the framework of the single-impurity Anderson Hamiltonian \cite{Anderson1}
 for two contrasting thermal limits ($T_{K}$ refers to the Kondo temperature): i) $T\gg T_{K}$, where the mean-field Hartree-Fock approach is applicable \cite{Uchoa1,Uchoa2}, and ii) $T\ll T_{K}$, a regime governed  by the formation of the Kondo cloud for which the role of strong correlation effects becomes crucial \cite{Kondo1,Kondo2,Kondo3}. For the latter, by adding an extra adatom to the host, an interesting effect emerges: the effective exchange coupling of localized spins exhibits the swap of its sign as the inter-adatoms separation
is changed. This is because the exchange between the localized spins is mediated by conducting electrons undergoing Friedel oscillations. Such mechanism forms the basis of the RKKY interaction, which in the case of graphene becomes strongly anisotropic \cite{RKKY1,RKKY2,RKKY3,RKKY4}.

In this Letter, employing the two-impurity Anderson Hamiltonian, we predict the formation of a multilevel structure in the local density of states (LDOS) of graphene and beats in the induced LDOS  in the vicinity of the Dirac points as the aftermath of the inter-adatoms correlations mediated by conducting electrons. To ensure the full absence of spin related phenomena provided by Kondo antiferromagnetic screening, we consider a non-magnetic host and work in the regime $T\gg T_{K}$. In doing so, we can safely focus on the regime where only charge fluctuations for the two adatoms placed far apart on the graphene sheet are relevant, cf.\,Fig.\,\ref{fig:Pic1}(a). Such fluctuations can be probed with an STM tip placed over a site of the sublattice $A$ or $B$ (see Figs.\ref{fig:Pic1}(b) and (c)) and result in the beating patterns in the induced LDOS to be discussed below.
Given the discrete nature of the graphene lattice we can measure the characteristic lengths by employing discrete indices, as follows: $m$ for inter-adatoms separations and $p$ designating the STM tip position (see Fig.\,\ref{fig:Pic1}(a)). We have found that to obtain the pronounced multilevel structure and distinct beating patterns, the constraint $m=2p$ for $p\gg1$ should be fulfilled \cite{obs2}. Additionally, we have found that the beats
are highly anisotropic, having different dependence along the zigzag and armchair directions. Our results point out that the LDOS is still sensitive to impurities separated by large distances, thus revealing that graphene is a suitable host for the observation of long-range interactions between adatoms.

\textit{The model.-}
To give a theoretical description of a such setup, the model based on the two-impurity Anderson
Hamiltonian treated in frameworks of Hubbard I approximation is developed. The Hamiltonian of the system reads:
\begin{align}
\mathcal{H}^{\text{2D}} & =-t\sum_{\bold k\sigma}[\phi({\bold k})a_{\bold k\sigma}^{\dagger}b_{\bold k\sigma}+\text{H.c.}]+\sum_{j\sigma}\mathcal{E}_{jd\sigma}d_{j\sigma}^{\dagger}d_{j\sigma}\nonumber \\
 & +\sum_{j}\mathcal{U}n_{d_{j}\uparrow}n_{d_{j}\downarrow}+[\sum_{j=1}^{2}\frac{\mathcal{V}_{j}}{\sqrt{\mathcal{N}}}\sum_{\bold k\sigma}e^{-i{\bold k}\cdot{\bold R}_{j}}(\phi^{*}({\bold k})a_{\bold k\sigma}^{\dagger}\nonumber \\
 & +\phi({\bold k})b_{\bold k\sigma}^{\dagger})d_{j\sigma}+\text{H.c.}],\label{eq:TIAM}
\end{align}
where $\phi({\bold k})=\sum_{i=1}^{3}e^{i{\bold k}\cdot{\bold\delta}_{i}},$ ${\bold\delta_{1}}=a{\bold e}_{x}$ and ${\bold\delta}_{2,3}=\frac{a}{2}(-{\bold e}_{x}\pm\sqrt{3}{\bold e}_{y})$
are the nearest neighbor vectors for adatoms placed at the center of the
hexagonal cells and $a\sim1.4$\,${\AA}$ is the side length. The surface electrons forming the host are described
by the operators $a_{\bold k\sigma}^{\dagger}$ ($a_{\bold k\sigma}$)
and $b_{\bold k\sigma}^{\dagger}$ ($b_{\bold k\sigma}$) for the
creation (annihilation) of an electron in a quantum state labeled
by the wave number ${\bold k}$ and spin $\sigma$ respectively in the
sublattices\textit{ $A$} and \textit{$B$}. For the adatoms,
$d_{j\sigma}^{\dagger}$ ($d_{j\sigma}$) creates (annihilates) an
electron with spin $\sigma$ in the state $\mathcal{E}_{jd\sigma}$,
with the index $j=1,2$. The third term in Eq.(\ref{eq:TIAM}) accounts
for the on-site Coulomb interaction $\mathcal{U}$, with $n_{d_{j}\sigma}=d_{j\sigma}^{\dagger}d_{j\sigma}$.
Finally, the last term mixes the host continuum of states of the graphene
and the discrete levels $\mathcal{E}_{jd\sigma}.$ This hybridization occurs
at the impurity sites via the coupling $\frac{\mathcal{V}_{j}}{\sqrt{\mathcal{N}}}e^{-i{\bold k}\cdot{\bold R}_{j}},$
with $\mathcal{N}$ being the total number of states, connected to the density of
states (DOS) per particle for graphene $\text{\ensuremath{\mathcal{D}}}_{0}=\frac{\Omega_{0}}{2\mathcal{N}\pi}\frac{\left|\mathcal{E}\right|}{(\hbar v_{F})^{2}}=\frac{\left|\mathcal{E}\right|}{D^{2}}$,
where $\Omega_{0}$ is the unit cell area, $v_{F}$ is the Fermi velocity and $D$ denotes the band-edge \cite{Uchoa1,Uchoa2}.

To determine
the  density of states (DOS) of the adatoms at the sites ${\bold R}_{j}$ in the host, we should calculate the Green's functions $\tilde{\mathcal{G}}_{d_{j\sigma}d_{l\sigma}}$
($j,l=1,2$), $\text{DOS}_{jj}^{\sigma}=-\frac{1}{\pi}{\tt Im}(\tilde{\mathcal{G}}_{d_{j\sigma}d_{j\sigma}})$.
To this end, the Hubbard I approximation can be used \cite{Hubbard,Hubbard1}. This approach provides reliable results away from the Kondo regime.
We start employing the equation-of-motion (EOM) method to a single particle retarded Green's function of an impurity in time domain
\begin{align}
\mathcal{G}_{d_{l\sigma}d_{j\sigma}} & =-\frac{i}{\hbar}\theta\left(t\right){\tt Tr}\{\varrho_{\text{2D}}[d_{l\sigma}\left(t\right),d_{j\sigma}^{\dagger}\left(0\right)]_{+}\},\label{eq:GF}
\end{align}
where $\theta\left(t\right)$ is the Heaviside function, $\varrho_{\text{2D}}$ is the density matrix of the system
described by the Hamiltonian {[}Eq. (\ref{eq:TIAM}){]} and $[\cdots,\cdots]_{+}$
is the anticommutator between operators taken in the Heisenberg picture. Performing elementary algebra one obtains in the energy domain:
\begin{align}
(\mathcal{E}^{+}-\mathcal{E}{}_{ld\sigma})\tilde{\mathcal{G}}_{d_{l\sigma}d_{j\sigma}} & =\delta_{lj}+\sum_{\tilde{l}}\Sigma{}_{\tilde{l}l}\tilde{\mathcal{G}}_{d_{\tilde{l}\sigma}d_{j\sigma}}\nonumber \\
 & +\mathcal{U}\tilde{\mathcal{G}}_{d_{l\sigma}n_{d_{l}\bar{\sigma}},d_{j\sigma}},\label{eq:s1}
\end{align}
where $\mathcal{E}^{+}=\mathcal{E}+i0^{+}$ and the self-energy given by $\Sigma_{\tilde{l}l(l\tilde{l})}({\bold d})=\frac{2\mathcal{V}_{l}\mathcal{V}_{\tilde{l}}}{N}\sum_{\bold k}e^{\mp i{\bold k}\cdot{\bold d}}\frac{\mathcal{E}^{+}|\phi({\bold k})|^{2}-t\text{Re}\big[\phi({\bold k})^{3}\big]}{\mathcal{E}^{+2}-t^{2}|\phi({\bold k})|^{2}},$
with ${\bold d}={\bold R}_{\tilde{l}}-{\bold R}_{l}$.

In the equation above, $\tilde{\mathcal{G}}_{d_{l\sigma}n_{d_{l}\bar{\sigma}},d_{j\sigma}}$
denotes a two particle Green's function composed by four fermionic operators,
obtained by Fourier transform of
\begin{align}
\mathcal{G}_{d_{l\sigma}n_{d_{l}\bar{\sigma}},d_{j\sigma}} & =-\frac{i}{\hbar}\theta\left(t\right){\tt Tr}\{\varrho_{\text{2D}}[d_{l\sigma}\left(t\right)n_{d_{l}\bar{\sigma}}\left(t\right),d_{j\sigma}^{\dagger}\left(0\right)]_{+}\},\label{eq:H_GF}
\end{align}
where $\bar{\sigma}=-\sigma$ and $n_{d_{l}\bar{\sigma}}=d_{l\bar{\sigma}}^{\dagger}d_{l\bar{\sigma}}$. In order to close
the system of the dynamic equations, we obtain the EOM for the Green's function given by Eq.(\ref{eq:H_GF}), which reads:
\begin{align}
(\mathcal{E}^{+}-\mathcal{E}_{ld\sigma}-\mathcal{U})\tilde{\mathcal{G}}_{d_{l\sigma}n_{d_{l}\bar{\sigma}},d_{j\sigma}} & =\delta_{lj}<n_{d_{l}\bar{\sigma}}>+\frac{\mathcal{V}_{j}}{\sqrt{\mathcal{N}}}\nonumber \\
\times\sum_{\bold ks}[-\phi_{s}|_{{\bold R}_{l}}({\bold k})\tilde{\mathcal{G}}_{c_{s\bold k\bar{\sigma}}^{\dagger}d_{l\bar{\sigma}}d_{l\sigma},d_{j\sigma}} & +\phi_{s}^{*}|_{{\bold R}_{l}}({\bold k})(\tilde{\mathcal{G}}_{c_{s\bold k\sigma}d_{l\bar{\sigma}}^{\dagger}d_{l\bar{\sigma}},d_{j\sigma}}\nonumber \\
+\tilde{\mathcal{G}}_{d_{l\bar{\sigma}}^{\dagger}c_{s\bold k\bar{\sigma}}d_{l\sigma},d_{j\sigma}}) & ],\label{eq:H_GF_2}
\end{align}
where the index $s=A,B$ marks a sublattice, $c_{A\bold k\sigma}=a_{\bold k\sigma}$
and $c_{B\bold k\sigma}=b_{\bold k\sigma}$, $\phi_{A}|_{{\bold R}_{l}}({\bold k})=e^{-i{\bold k}\cdot{\bold R}_{l}}\phi^{*}({\bold k})$
and $\phi_{B}|_{{\bold R}_{l}}({\bold k})=e^{-i{\bold k}\cdot{\bold R}_{l}}\phi({\bold k})$,
expressed in terms of new Green's functions of the same order of $\tilde{\mathcal{G}}_{d_{l\sigma}n_{d_{l}\bar{\sigma}},d_{j\sigma}}$
and the occupation number
\begin{equation}
<n_{d_{l}\bar{\sigma}}>=-\frac{1}{\pi}\int_{-D}^{+D}n_{F}(\mathcal{E}){\tt Im}(\tilde{\mathcal{G}}_{d_{l{\bar{\sigma}}}d_{l{\bar{\sigma}}}})d\mathcal{E},\label{eq:nb}
\end{equation}where $n_{F}(\mathcal{E})$ is the Fermi-Dirac distribution. By employing
the Hubbard I approximation, we decouple the Green's
functions in the right-hand side of Eq.(\ref{eq:H_GF_2}), as follows:
$\tilde{\mathcal{G}}_{c_{s\bold k\bar{\sigma}}^{\dagger}d_{l\bar{\sigma}}d_{l\sigma},d_{j\sigma}}\simeq<c_{s\bold k\bar{\sigma}}^{\dagger}d_{l\bar{\sigma}}>\tilde{\mathcal{G}}_{d_{l\sigma}d_{j\sigma}}$
and $\tilde{\mathcal{G}}_{d_{l\bar{\sigma}}^{\dagger}c_{s\bold k\bar{\sigma}}d_{l\sigma},d_{j\sigma}}\simeq<c_{s\bold k\bar{\sigma}}^{\dagger}d_{l\bar{\sigma}}>\tilde{\mathcal{G}}_{d_{l\sigma}d_{j\sigma}}$,
where we have used $\sum_{\bold ks}\phi({\bold k})e^{-i{\bold k}\cdot{\bold R}_{l}}=\sum_{\bold ks}\phi^{*}({\bold k})e^{i{\bold k}\cdot{\bold R}_{l}}$.
As a result, we find
\begin{eqnarray}
(\mathcal{E}^{+}-\mathcal{E}_{ld\sigma}-\mathcal{U})\tilde{\mathcal{G}}_{d_{l\sigma}n_{d_{l}\bar{\sigma}},d_{j\sigma}} & = & \delta_{lj}<n_{d_{l}\bar{\sigma}}>\nonumber \\
+\frac{\mathcal{V}_{j}}{\sqrt{\mathcal{N}}}\sum_{\bold ks} & \phi_{s}^{*}|_{\bold{R}_{l}}({\bold k}) & \tilde{\mathcal{G}}_{c_{s\bold k\sigma}d_{l\bar{\sigma}}^{\dagger}d_{l\bar{\sigma}},d_{j\sigma}}.\nonumber \\
\label{eq:H_GF_3-1}
\end{eqnarray}
To complete the calculation, we need to determine $\tilde{\mathcal{G}}_{c_{s\bold k\sigma}d_{l\bar{\sigma}}^{\dagger}d_{l\bar{\sigma}},d_{j\sigma}}$.
Once again, employing the EOM approach for $\tilde{\mathcal{G}}_{c_{s\bold k\sigma}d_{l\bar{\sigma}}^{\dagger}d_{l\bar{\sigma}},d_{j\sigma}}$,
we obtain
\begin{align}
\mathcal{E}^{+}\tilde{\mathcal{G}}_{c_{s\bold k\sigma}d_{l\bar{\sigma}}^{\dagger}d_{l\bar{\sigma}},d_{j\sigma}} & =-t\phi_{\bar{s}}|_{\bold{R}_{l}\bold{=0}}({\bold k})\tilde{\mathcal{G}}_{c_{\bar{s}\bold k\sigma}d_{l\bar{\sigma}}^{\dagger}d_{l\bar{\sigma}},d_{j\sigma}}\nonumber \\
+ & \sum_{\bold q\tilde{s}}\frac{\mathcal{V}_{l}}{\sqrt{\mathcal{N}}}\phi_{\tilde{s}}^{*}|_{{\bold R}_{l}}({\bold q})\tilde{\mathcal{G}}_{c_{s\bold k\sigma}d_{l\bar{\sigma}}^{\dagger}c_{\tilde{s}\bold q\bar{\sigma}},d_{j\sigma}}\nonumber \\
+ & \sum_{\tilde{j}}\frac{\mathcal{V}_{\tilde{j}}}{\sqrt{\mathcal{N}}}\phi_{s}|_{{\bold R}_{\tilde{j}}}({\bold k})\tilde{\mathcal{G}}_{d_{\tilde{j}\sigma}n_{d_{l}\bar{\sigma}},d_{j\sigma}}\nonumber \\
- & \sum_{\bold q\tilde{s}}\frac{\mathcal{V}_{l}}{\sqrt{\mathcal{N}}}\phi_{\tilde{s}}|_{{\bold R}_{l}}({\bold q})\tilde{\mathcal{G}}_{c_{\tilde{s}\bold q\bar{\sigma}}^{\dagger}d_{l\bar{\sigma}}c_{s\bold k\sigma},d_{j\sigma}},\nonumber \\
\label{eq:H_GF_4}
\end{align}
where $\bar{s}=A,B$ respectively for $s=B,A$ as labels to correlate simultaneously distinct sublattices, while $\tilde{s}=A,B$ runs arbitrarily. For the
sake of simplicity, we take the limit $\mathcal{U}\rightarrow\infty$
and continue with the Hubbard I scheme by making $\tilde{\mathcal{G}}_{c_{s\bold k\sigma}d_{l\bar{\sigma}}^{\dagger}c_{\tilde{s}\bold q\bar{\sigma}},d_{j\sigma}}\simeq\left\langle d_{l\bar{\sigma}}^{\dagger}c_{\tilde{s}\bold q\bar{\sigma}}\right\rangle \tilde{\mathcal{G}}_{c_{s\bold k\sigma}d_{j\sigma}},$
$\tilde{\mathcal{G}}_{c_{\tilde{s}\bold q\bar{\sigma}}^{\dagger}d_{l\bar{\sigma}}c_{s\bold k\sigma},d_{j\sigma}}\simeq\left\langle d_{l\bar{\sigma}}^{\dagger}c_{\tilde{s}\bold q\bar{\sigma}}\right\rangle \tilde{\mathcal{G}}_{c_{s\bold k\sigma}d_{j\sigma}}$
and $\tilde{\mathcal{G}}_{d_{\tilde{j}\sigma}n_{d_{l}\bar{\sigma}},d_{j\sigma}}\simeq\left\langle n_{d_{l}\bar{\sigma}}\right\rangle \tilde{\mathcal{G}}_{d_{\tilde{j}\sigma}d_{j\sigma}}$
in Eq.(\ref{eq:H_GF_4}), which in combination with Eqs.\,(\ref{eq:s1})
and (\ref{eq:H_GF_3-1}) results in
\begin{equation}
\tilde{\mathcal{G}}_{d_{j\sigma}d_{j\sigma}}=\frac{1-<n_{d_{j}\bar{\sigma}}>}{\mathcal{E}-\mathcal{E}_{jd\sigma}-{{\tilde{\Sigma}}^{\sigma}}_{jj}},\label{eq:pass3-1}
\end{equation}
where
\begin{equation}
{{\tilde{\Sigma}}^{\sigma}}_{jj}=\Sigma{}_{jj}+\lambda_{j\bar{j}}^{\bar{\sigma}}\frac{\Sigma_{j\bar{j}}({\bold d})\Sigma_{\bar{j}j}({\bold d})}{\mathcal{E}-\mathcal{E}_{\bar{j}d\sigma}-\Sigma_{\bar{j}\bar{j}}}\label{eq:TSE}
\end{equation}
is the total self-energy, $\lambda_{j\bar{j}}^{\bar{\sigma}}=(1-\left\langle n_{d_{j}\bar{\sigma}}\right\rangle )(1-\left\langle n_{d_{\bar{j}}\bar{\sigma}}\right\rangle )$, with $\bar{j}=1,2$ respectively for $j=2,1$ as indexes to correlate distinct adatoms and

\begin{eqnarray}
\tilde{\mathcal{G}}_{d_{j\sigma}d_{\bar{j}\sigma}} & = & (1-<n_{d_{j}\bar{\sigma}}>)\frac{\Sigma_{\bar{j}j}({\bold d})\tilde{\mathcal{G}}_{d_{\bar{j}\sigma}d_{\bar{j}\sigma}}}{\mathcal{E}-\mathcal{E}_{jd\sigma}-\Sigma_{jj}}\label{eq:G12}
\end{eqnarray}
accounting for the crossed Green's function.

In the vicinity of the Dirac points $\bold K_{\pm}=2\pi/3a(1,\pm1/\sqrt{3})$
we obtain $t|\phi({\bold k})|=\hbar v_{F}k$ and for adatoms equally coupled to the graphene host $(\mathcal{V}_{1}=\mathcal{V}_{2}=\mathcal{V})$,
we determine the following self-energies \cite{Uchoa2},
\begin{align}
\Sigma{}_{11} & =\Sigma{}_{22}=-2\frac{\mathcal{V}^{2}}{D^{2}}[\frac{\mathcal{E}}{t^{2}}(D^{2}+\mathcal{E}^{2}\ln\Big|\frac{D^{2}-\mathcal{E}^{2}}{\mathcal{E}^{2}}\Big|)\nonumber \\
 & +i\pi\frac{|\mathcal{E}|^{3}}{t^{2}}\theta(D-\mathcal{E})]\label{eq:SE1}
\end{align}
and
\begin{align}
\Sigma{}_{12(21)}({\bold d}) & =(e^{\mp i{\bold K_{+}}\cdot{\bold d}}+e^{\mp i{\bold K_{-}}\cdot{\bold d}})\frac{\pi}{i}\frac{\mathcal{V}^{2}}{D^{2}}\frac{|\mathcal{E}|^{3}}{t^{2}}H_{0}^{(1)}\left(\frac{\mathcal{E}|{\bold d}|}{\hbar v_{F}}\right),\nonumber \\
\label{eq:SE2}
\end{align}where $H_0^{(1)}$ stands for the \textit{zeroth}-order
Hankel function of the first kind. The expression is valid in the range of small energies
where $|\mathcal{E}|\ll D$ and for distant adatoms characterized by the ratio $|\frac{\mathcal{E}|{\bold d}|}{\hbar v_{F}}|\gg1$
\cite{Friedel1}. To obtain the host LDOS probed by the STM tip of Fig.\,\ref{fig:Pic1} we introduce the retarded
Green's function in time coordinate, which reads
\begin{align}
\mathcal{G}_{\sigma}({\bold r}_{s},t) & =-\frac{i}{\hbar}\theta\left(t\right){\tt Tr}\{\varrho_{\text{2D}}[\tilde{\Psi}_{\sigma}({\bold r}_{s},t),\tilde{\Psi}_{\sigma}^{\dagger}({\bold r}_{s},0)]_{+}\}\nonumber \\
\label{eq:PSI_R}
\end{align}
with
\begin{equation}
\tilde{\Psi}_{\sigma}({\bold r}_{s})=\frac{1}{\sqrt{\mathcal{N}}}\sum_{\bold k}e^{i{\bold k}\cdot{\bold r}_{s}}c_{s\bold k\sigma}\label{eq:PSI_R-1-1}
\end{equation}
as the field operator accounting for the quantum state of the graphene
site placed right beneath the tip, with $s=A,B$ designating the sublattices
of the system, thus resulting in $c_{A\bold k\sigma}=a_{\bold k\sigma}$
and $c_{B\bold k\sigma}=b_{\bold k\sigma}.$ Therefore, the LDOS at
a site ${\bold r}_{s}$ of the host can be obtained as
\begin{equation}
\text{{LDOS}}({\bold r}_{s})=-\frac{1}{\pi}{\tt Im}[\tilde{\mathcal{G}}_{\sigma}(\varepsilon^{+},{\bold r}_{s})],\label{eq:FM_LDOS}
\end{equation}
where $\tilde{\mathcal{G}}_{\sigma}(\varepsilon^{+},{\bold r}_{s})$
is the time Fourier transform of $\mathcal{G}_{\sigma}(t,{\bold r}_{s}).$ Then by applying the equation of motion (EOM) on Eq. (\ref{eq:PSI_R}),
one can show that $\text{{LDOS}}({\bold r}_{s})=\text{\ensuremath{\mathcal{D}}}_{0}+\Delta\text{{LDOS}}({\bold r}_{s})=\text{\ensuremath{\mathcal{D}}}_{0}+\sum_{jl}\Delta\text{{LDOS}}_{jl}({\bold r}_{s})$,
with

\begin{align}
\Delta\text{{LDOS}}({\bold r}_{s})_{jl} & =-(\pi\mathcal{V}^{2}\text{\ensuremath{\mathcal{D}}}_{0}^{2}){\tt Im}[(q_{jr}-i\mathcal{F}_{jr})\tilde{\mathcal{G}}_{d_{j\sigma}d_{l\sigma}}\nonumber \\
 & \times(q_{rl}-i\mathcal{F}_{rl})]\label{eq:LDOSp1}
\end{align}describing the renormalization of the LDOS by the adatoms. It depends on
the graphene site ${\bold r}_{s}$ as outlined in Figs.\,\ref{fig:Pic1}(b) and (c),
where $s=A,B$ denotes the type of sublattices of the system, $q_{jr}=\frac{1}{\pi\mathcal{V}^{2}\text{\ensuremath{\mathcal{D}}}_{0}}\text{{\tt Re}}\Sigma_{jr}({\bold d}_{j})$
describes the Fano parameter of interference \cite{Fano} and $\mathcal{F}_{jr}=-\frac{1}{\pi\mathcal{V}^{2}\text{\ensuremath{\mathcal{D}}}_{0}}\text{{\tt Im}}\Sigma_{jr}({\bold d}_{j})$ gives rise to the Friedel oscillations in the graphene sheet, where ${\bold d}_{j}={\bold R}_{j}-{\bold r}_{s}$ and ${\bold r}_{s}\neq{\bold R}_{j}$. The $\text{{LDOS}}({\bold r}_{s})$ is spin-independent since graphene is not ferromagnetic. As a result of substituting Eqs. (\ref{eq:pass3-1}) and (\ref{eq:TSE}) into Eq. (\ref{eq:LDOSp1}), we show that the diagonal term $l=j$ leads to

\begin{align}
\Delta\text{{LDOS}}_{jj}({\bold r}_{s}) & =a({\bold d}_{j})\frac{|\frac{q_{jr}}{\mathcal{F}_{jr}}|^{2}-1+2\xi_{j}{\tt Re}(\frac{q_{jr}}{\mathcal{F}_{jr}})}{\xi_{j}^{2}+1}\nonumber \\
\label{eq:LDOSp1-1-1}
\end{align}as the contribution arising from the \textit{jth} adatom obeying the Fano-like expression of Ref.\,{[}\onlinecite{FanoR}{]}, in which $a({\bold d}_{j})=(1-<n_{d_{j}\bar{\sigma}}>)\frac{\pi\mathcal{V}^{2}\text{\ensuremath{\mathcal{D}}}_{0}^{2}}{\Delta_{jj}}|\mathcal{F}_{jr}|^{2},$ $\xi_{j}=\frac{\mathcal{E}-(\mathcal{E}_{jd\sigma}+{\tt Re}{{\tilde{\Sigma}}^{\sigma}}_{jj})}{\Delta_{jj}}$ and $\Delta_{jj}=-{\tt Im}{{\tilde{\Sigma}}^{\sigma}}_{jj}.$ It is worth mentioning that the couple of Eqs. (\ref{eq:LDOSp1}) and (\ref{eq:LDOSp1-1-1}) constitutes the main analytical
findings of this work: \textit{for two adatoms far apart, the LDOS signal captured by the STM probe is mainly ruled by the interference between two scattered waves shaped by Fano-like forms following Eq. (\ref{eq:LDOSp1-1-1})}.

\textit{Results and Discussion.-}
The system of the dynamical equations we have obtained allows us to investigate the effect of a pair of correlated impurities on the LDOS of graphene host. Our approach is valid for $T\gg T_{K}$ and within a range of temperatures where we can safely define the Heaviside step function in Eq.\,(\ref{eq:nb}) for the Fermi-Dirac distribution $n_{F}(\mathcal{E})$. This assumption was previously
considered in Ref.\,{[}\onlinecite{Seridonio1}{]}.
The relevant parameter of the model which strongly affects the beating pattern is the Fermi velocity in the Dirac point, $v_{F}=\frac{3}{2}\frac{at}{\hbar}$ \cite{Peres,CNeto1}. For an individual graphene sheet in vacuum it is equal approximately to $c/300$, where $c$ denotes the speed of light. Note, however, that recently it was proposed that the tuning of the Fermi velocity can be achieved experimentally by changing the dielectric constant in the substrate of the graphene sheet \cite{TFV}. In our calculations we have adopted $\mathcal{V}=t=0.1D$ (it corresponds to $v_{F}\sim\frac{c}{1200}$) and $\mathcal{E}_{1d\sigma}=\mathcal{E}_{2d\sigma}=-0.09D,$ with $D=7$\,eV as the graphene band-edge \cite{Uchoa1,Uchoa2}, ${\bold R}_{1}={\bold0}$ and ${\bold R}_{2}={\bold d}$ to set the displacement of the second adatom with respect to the first by following ${\bold d}=\sqrt{3}ma{\bold e}_{y}$ and ${\bold d}=3ma{\bold e}_{x}$, respectively for the zigzag and armchair directions, with $m=1,2,3,\ldots$ as an integer number.
In the case of the displacement of the STM tip along the armchair direction, we have found ${\bold r}_{A}=(1+3p)a{\bold e}_{x}$ for sites in the
sublattice $A$ and ${\bold r}_{B}=(2+3p)a{\bold e}_{x}$ for those in the sublattice $B$. Similar analysis for the zigzag direction leads to ${\bold r}_{A}=\frac{\sqrt{3}}{2}(1+2p)a{\bold e}_{y}$ and ${\bold r}_{B}=\frac{\sqrt{3}}{2}(2p)a{\bold e}_{y}$, respectively for sublattices $A$ and $B$. In both directions, we have the index $p=0,1,2,\ldots.$ We have found that by imposing the constraint $m=2p$ for $p\gg1,$ the presence of two extremely distant adatoms still affects the graphene LDOS giving rise to an anisotropic multilevel structure and beating patterns. In our analysis, we have used the value $p=35.$ Such a choice leads to ${|\bold d|}\sim294$\,${\AA},$ $|{\bold r}_{A}|\sim148$\,${\AA},$ $|{\bold r}_{B}|\sim150$\,${\AA}$ and ${|\bold d|}\sim167$\,${\AA}$, $|{\bold r}_{A}|\sim60$\,${\AA},$ $|{\bold r}_{B}|\sim59.5$\,${\AA},$ respectively for the armchair and zigzag directions.

\begin{figure}
\includegraphics[width=0.4\textwidth,height=0.179\textheight]{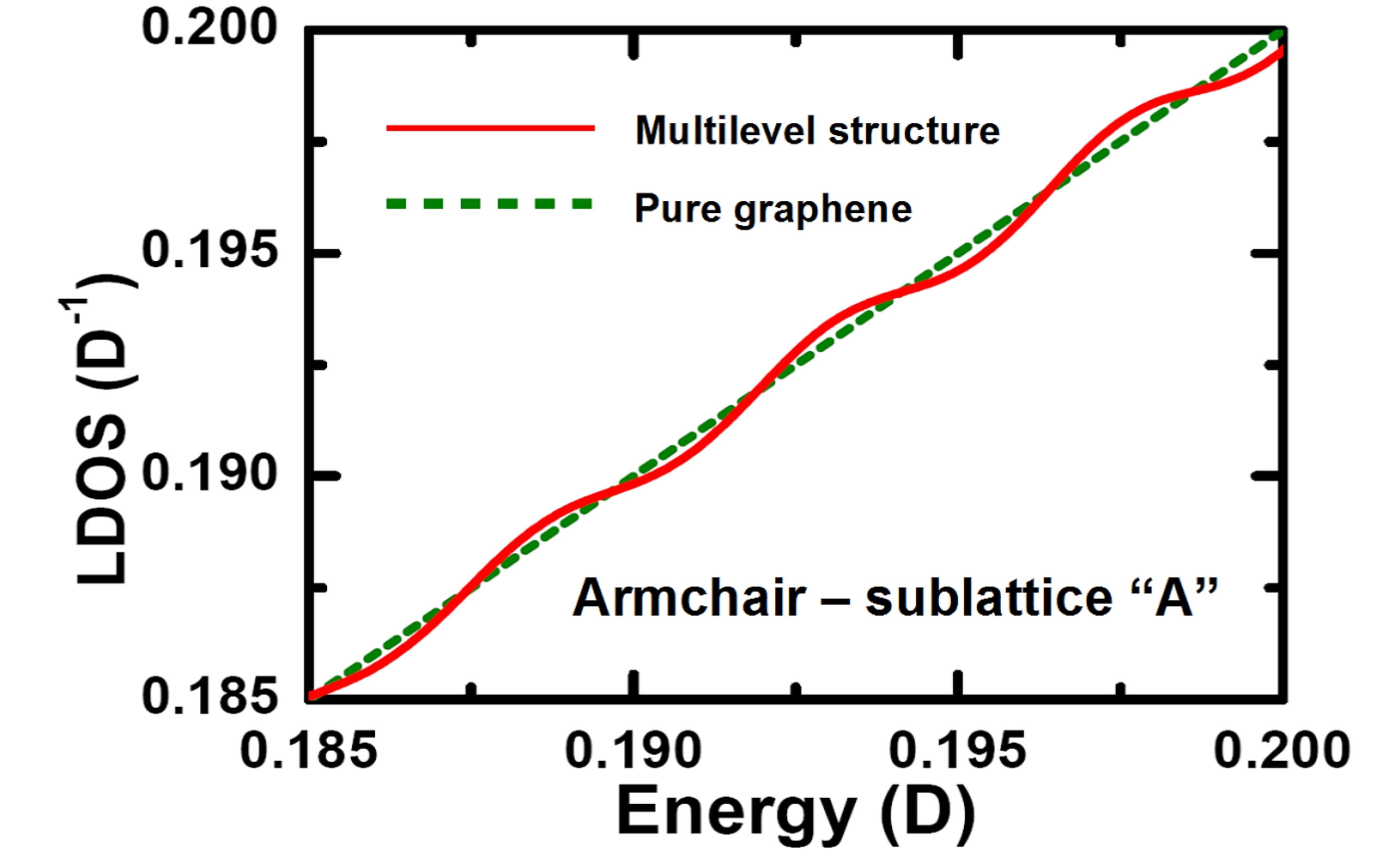}
\caption{\label{fig:Pic2} (Color online) $\text{{LDOS}}({\bold r}_{A})$ as a function of energy for the armchair direction: a multilevel structure emerges.}
\end{figure}

\begin{figure}
\includegraphics[width=0.48\textwidth,height=0.149\textheight]{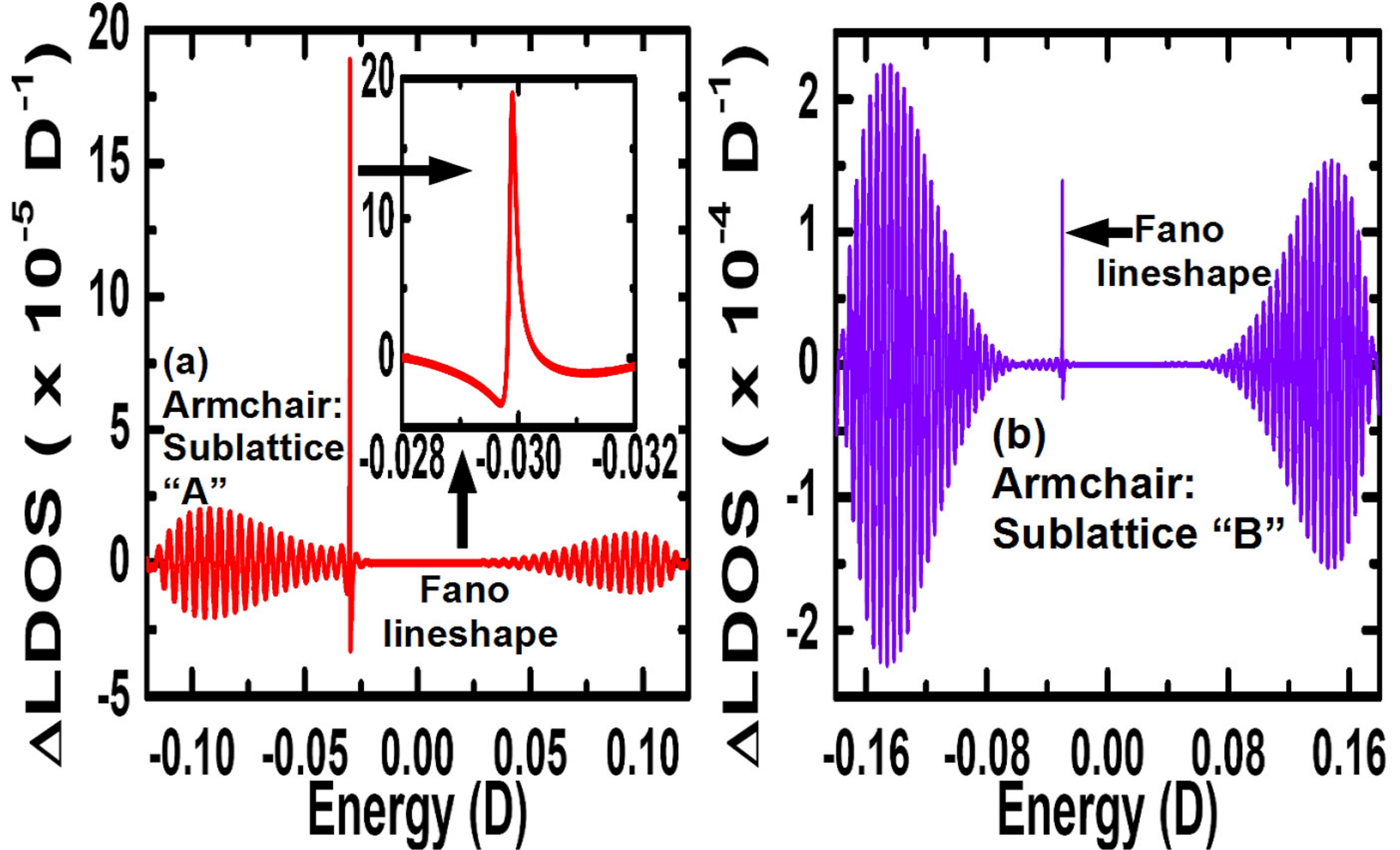}
\caption{\label{fig:Pic3}(Color online) Beating pattern in the LDOS corresponding to the armchair placement of the impurities for sublattices A (panel (a)) and B (panel (b)). Note the presence of a sharp Fano lineshape corresponding to the presence of the localized states (inset of the panel (a)).}
\end{figure}

\begin{figure}
\includegraphics[width=0.48\textwidth,height=0.149\textheight]{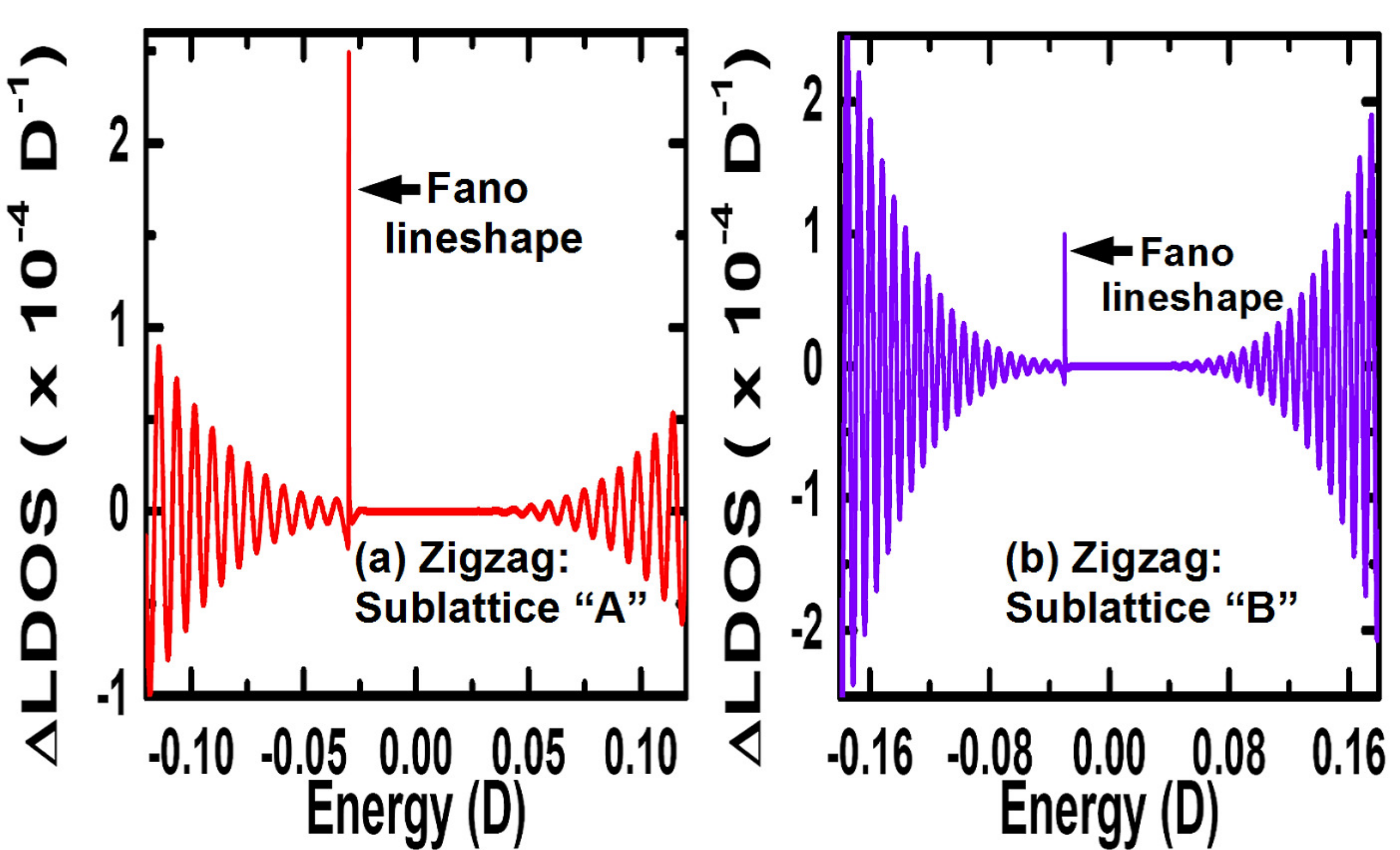}
\caption{\label{fig:Pic4}(Color online) (a) $\Delta\text{{LDOS}}({\bold r}_{A})$ as a function of energy for the zigzag placement of the impurities for sublattices A (panel (a)) and B (panel (b)). Note that although the multilevel structure is clearly seen the beats are absent.}
\end{figure}

In Fig.\ref{fig:Pic2}, we show the behavior of the $\text{{LDOS}}({\bold r}_{A})=\text{\ensuremath{\mathcal{D}}}_{0}+\Delta\text{{LDOS}}({\bold r}_{A})$
for the armchair direction as a function of energy $\mathcal{E}$.
Above and below (not shown) $\bold K_{\pm}$ (Fermi level), the total LDOS presents a resolved multilevel structure, cf.\,Fig.\,\ref{fig:Pic2}. The LDOS for pure graphene is represented by the dotted-green line. The corresponding
profile for sublattice $B$ as well as those in the zigzag direction
are very similar to Fig.\,\ref{fig:Pic2} and are not presented here. Interestingly enough, the noise within the experimental data of the differential conductance reported for the epitaxial graphene embedding atomic defects is reminiscent of the multilevel structure obtained theoretically in the frame of this work (see panels (j) to (m) of Fig.\,$3$ in Ref.\,{[}\onlinecite{SciI}{]}). Particularly for this system, the intervalley scattering is recognized by the authors as the underlying mechanism for this feature. In which concerns the setup of Fig.\ref{fig:Pic1}, the multilevel behavior lies on the Fano interference assisted by a couple of adatoms as the expression for $\text{{LDOS}}({\bold r}_{s})$ and Eq. (\ref{eq:LDOSp1-1-1}) ensures.

Thus by subtracting the background $\text{\ensuremath{\mathcal{D}}}_{0}$ from $\text{{LDOS}}({\bold r}_{A}),$
a beating pattern composed by a pair of wave packets is revealed in
$\Delta\text{{LDOS}}({\bold r}_{A})$ as shown in Fig.\,\ref{fig:Pic3}(a)
for the armchair direction. For sublattice $B$, the beating pattern
of $\Delta\text{{LDOS}}({\bold r}_{B})$
exhibits even more pronounced amplitude as shown in Fig.\,\ref{fig:Pic3}(b). Despite of the moderate amplitude within $\Delta\text{{LDOS}}({\bold r}_{s})$ revealed by the simulations, we stress that the differential conductance $\Delta G\sim2\frac{e^{2}}{h}\Gamma_{\text{{tip}}}\Delta\text{{LDOS}}({\bold r}_{s})$ \cite{Uchoa2} is indeed the quantity  measured by the STM probe, where $\Gamma_{\text{{tip}}}$ is the graphene-tip coupling. By moving vertically the tip towards the graphene sheet, such a coupling increases and leads to the enhancement of the signal, thus allowing its experimental detection. Additionally, we point out that the unpronounced magnitude of the LDOS reported here attests the signature of a long-range perturbation induced by defects as that previously observed in a similar system composed by graphite and adsorbed molecules \cite{SciII}.

In both sublattices of the system considered in Fig.\,\ref{fig:Pic1}, the localized states $\mathcal{E}_{1d\sigma}=\mathcal{E}_{2d\sigma}=-0.09D$ of
the adatoms are characterized by Fano lineshapes (see inset of Fig.\,\ref{fig:Pic3}(a)). Remarkably, the position of localized levels becomes renormalized and are given by $\mathcal{\tilde{E}}_{1d\sigma}=\mathcal{\tilde{E}}_{2d\sigma}=-0.03D$ (inset of Fig.\,\ref{fig:Pic3}(a)), due to the anomalous shifting $\frac{\mathcal{E}^{3}}{t^{2}}$ within $\Sigma{}_{11(22)}({\bold d})$ \cite{Uchoa2}.
As for the zigzag direction, although the multilevel structure is clearly observed, beating patterns are absent in $\Delta\text{{LDOS}}({\bold r}_{s})$, see Figs.\,\ref{fig:Pic4}(a) and (b). Such observations suggest that the formation of beats in the $\Delta\text{{LDOS}}({\bold r}_{s})$ of graphene is highly anisotropic. These phenomena arise from the interplay between the anomalous broadening $\frac{|\mathcal{E}|^{3}}{t^{2}}$ of $\Delta\text{{LDOS}}({\bold r}_{s})$ and the oscillations within $q_{jr}$ and  $\mathcal{F}_{jr},$ provided by Fano and Friedel effects respectively, which are enhanced by such a broadening.

Moreover, in the domain of large inter-adatoms separations as considered here ($m=2p$ and $p\gg1$), the damping nature of the Friedel oscillations prevails in the LDOS and the direct terms $\Delta\text{{LDOS}}_{jj}({\bold r}_{s})$
overcome the crossed $\Delta\text{{LDOS}}({\bold r}_{s})_{jl}$ when $j \neq l$ within Eq. (\ref{eq:LDOSp1}), thus resulting in patterns for $\Delta\text{{LDOS}}({\bold r}_{s})$
dictated by the superpositions of waves shaped by the Fano-like expression of Eq. (\ref{eq:LDOSp1-1-1}). Thereby, depending on
the direction in graphene, such waves can yield beating patterns and a multilevel structure as the aftermath of the interference between $\Delta\text{{LDOS}}_{11}({\bold r}_{s})$
and $\Delta\text{{LDOS}}_{22}({\bold r}_{s}),$ since $\Delta\text{{LDOS}}_{jj}({\bold r}_{s})$ encloses information on the electronic
wave of the host scattered by the \textit{jth} adatom.

\textit{Conclusions.-}
In summary, we have proposed an experimentally friendly setup based on monolayer graphene in which the long-range correlations between distantly placed adatoms can be detected.  We predict that the interplay between Fano and Friedel terms nearby the Dirac points leads to a multilevel structure and anisotropic beating patterns in the LDOS, which can be detected by STM measurements.

\textit{Acknowledgments.-} This work was supported by the agencies CNPq, CAPES, PROPG-PROPe/UNESP, FAPEMIG, FP7 IRSES projects
SPINMET and QOCaN. A.\,C.\,Seridonio thanks the University of Iceland and the Nanyang Technological University at Singapore for hospitality.


\begin{thebibliography}{10}
\bibitem{Novoselov}K. S. Novoselov, Rev. Mod. Phys. \textbf{83},
837 (2011).

\bibitem{Peres}N. M. R. Peres, Rev. Mod. Phys. \textbf{82}, 2673
(2010).

\bibitem{CNeto1}A. H. Castro Neto \textit{et al.},
Rev. Mod. Phys. \textbf{81}, 109 (2009).

\bibitem{Katayama} S. Katayama \textit{et al.}, J. Phys. Soc. Jpn. \textbf{75},  054705 (2006).

\bibitem{Hasan2010} M. Z. Hasan and C. L. Kane, Rev. Mod. Phys. \textbf{82}, 3045 (2010).

\bibitem{Eelbo1}T. Eelbo \textit{et al.},
Phys. Rev. B \textbf{87}, 205443 (2013).

\bibitem{Eelbo2}T. Eelbo \textit{et al.},
Phys. Rev. Lett. \textbf{110}, 136804 (2013).

\bibitem{abInitio1}X. Liu \textit{et al.},
Phys. Rev. B \textbf{83}, 235411 (2011).

\bibitem{STMreview}M. Ternes, A. J. Heinrich, and W.-D. Schneider,
J. Phys.: Condens. Matter \textbf{21}, 053001 (2009).

\bibitem{Friedel1}A. Bácsi, and A. Virosztek, Phys. Rev. B \textbf{82},
193405 (2010).

\bibitem{Friedel2}C. Bena, Phys. Rev. B \textbf{79}, 125427 (2009).

\bibitem{Anderson1}P. W. Anderson, Phys. Rev. \textbf{124}, 41 (1961).

\bibitem{Uchoa1}B. Uchoa \textit{et al.},
Phys. Rev. Lett. \textbf{101}, 026805 (2008).

\bibitem{Uchoa2}B. Uchoa \textit{et al.},
Phys. Rev. Lett. \textbf{103}, 206804 (2009).

\bibitem{Kondo1}Z. G. Zhu, and J. Berakdar, Phys. Rev. B \textbf{84},
165105 (2011).

\bibitem{Kondo2}B. Uchoa, T. G. Rappoport, and A. H. Castro Neto,
Phys. Rev. Lett. \textbf{106}, 016801 (2011).

\bibitem{Kondo3}L. Lin \textit{et al.}, New J. Phys. \textbf{15},
053018 (2013).

\bibitem{RKKY1}M. Sherafati, and S. Satpathy, Phys. Rev. B \textbf{83},
165425 (2011).

\bibitem{RKKY2}F. Parhizgar \textit{et al.},
Phys. Rev. B \textbf{87}, 125402 (2013).

\bibitem{RKKY3}P. D. Gorman \textit{et al.},
Phys. Rev. B \textbf{88}, 085405 (2013).

\bibitem{RKKY4}E. Kogan, Phys. Rev. B \textbf{84}, 115119 (2013).

\bibitem{obs2}The constraint  $m=2p$ for $p\gg1$ was determined numerically.

\bibitem{Hubbard}J. Hubbard, Proc. R. Soc. Lond. A, \textbf{281}, 401 (1964).

\bibitem{Hubbard1}H. Haug, and A. P. Jauho, Quantum Kinetics in Transport
and Optics of Semiconductors, Springer series in Solid-State Sciences
123 (Springer, New York, 1996).

\bibitem{Fano}A. E. Miroshnichenko, S. Flach, and Y. S. Kivshar,
Rev. Mod. Phys. \textbf{82},2257 (2010).

\bibitem{FanoR}C.-Y. Lin, A. H. Castro Neto, and B. A. Jones, Phys. Rev. Lett. \textbf{97}, 156102 (2006).

\bibitem{Seridonio1}A. C. Seridonio \textit{et al.},
Phys. Rev. B \textbf{88},
195122 (2013).

\bibitem{TFV}C. Hwang \textit{et al.}, Sci. Rep. \textbf{2}, 590 (2012); D. A. Siegel \textit{et al.}, Phys. Rev. Lett. \textbf{110}, 146802 (2013).

\bibitem{SciI}G. M. Rutter \textit{et al.} Science \textbf{317}, 219 (2007).

\bibitem{SciII}H. A. Mizes and J. S. Foster, Science \textbf{244}, 599 (1989).\end{thebibliography}
\end{document}